\documentclass[aps,prd,twocolumn,superscriptaddress,nofootinbib]{revtex4-2}

\usepackage{natbib}
\usepackage[T1]{fontenc}
\usepackage{amsmath}
\usepackage{amsfonts}
\usepackage{amssymb}
\usepackage{graphicx}
\usepackage{hyperref}
\usepackage{subfig}
\usepackage{float}

\newcommand{\Hebar}{{}^{3}\bar{\rm{He}}}
\newcommand{\Dbar}{\bar{\rm{D}}}

\begin{document}

\title{Late-forming black holes and the antiproton, gamma-ray, and anti-helium excesses}

\author{Mrunal Korwar}
\affiliation{Department of Physics, University of California, Berkeley, CA 94720, USA}
\affiliation{Department of Physics and Astronomy, University of Kentucky, Lexington, KY 40506, USA}

\author{Stefano Profumo}
\affiliation{Department of Physics and Santa Cruz Institute for Particle Physics, University of California, Santa Cruz, CA 95064, USA}

\thanks{Email: mkorwar@berkeley.edu}
\thanks{Email: profumo@ucsc.edu}

\begin{abstract}
    Black holes long-lived enough to be the dark matter have temperatures below the MeV. Since Hawking evaporation is a quasi-thermal process, no GeV emission is predicted to be produced by black holes if they are part, or all, of the cosmological dark matter. However, black holes could be ``spawned'' at late times with masses that correspond to short lifetimes, and as such be significantly hotter and produce particles well in excess of the GeV. Here we investigate if such late-forming black
holes could, at once, explain the tantalizing excesses found in the gamma radiation from the Galactic center, in the flux of cosmic-ray antiproton, and in the few tentative antihelium events reported by the anti-matter spectrometer AMS-02. We find that late-forming black holes cannot simultaneously explain all these excesses. We additionally compute accurate predictions for the anti-deuteron, high-energy neutrino, and positron fluxes if this scenario is realized in nature. We find that while the neutrino and positron fluxes are too small compared to the expected background, a significant number of anti-deuteron events is expected both at AMS-02 and at the future General AntiParticle Spectrometer (GAPS).
\end{abstract}

\preprint{N3AS-24-012}
\maketitle

\section{Introduction}
\label{sec:intro}

The nature of the cosmological dark matter (DM), posited in the standard cosmological model, remains elusive (for a constantly updated review, see Chapter 27 of \cite{ParticleDataGroup:2022pth}). The paradigm of weakly interacting particles (WIMPs), one where the DM is a weak-scale particle coupled, however weakly, typically in fact via weak interactions, to Standard Model particles, was considered for a long time as one of the most promising frameworks; as a result, a vast program of direct, indirect, and collider searches for WIMPs was undertaken, yielding, however, mostly negative results (see e.g. \cite{Arcadi:2017kky, Arcadi:2024ukq} for an extensive review of the WIMP paradigm). Nevertheless, a few tantalizing anomalies, which we refer to as ``excesses'', have been reported over the years that could potentially be ascribed to new physics effects sourced by a DM particle (for a pedagogical and complete review of indirect DM detection and the excesses described below see especially Ref.~\cite{Slatyer:2021qgc}).

One of the most persistent and intriguing example of such ``excesses'' is the Galactic center excess, which refers to an excess at a few GeV energies in the Fermi Large Area Telescope (LAT) \cite{Fermi-LAT:2009ihh} data over the expected diffuse gamma radiation in the inner Galaxy, originally discovered in Ref.~\cite{Goodenough:2009gk, Hooper:2010mq}. A significant debate followed, spurred by the potential contributions of unresolved point sources, a possible background mismodeling due to the lack of time-dependent effects or the localization of the cosmic-ray sources, or the very templates used to build a model of the diffuse emission, among other possibilities (we refer the Reader for details to the comprehensive chapter 6 of Ref.~\cite{Slatyer:2021qgc}). Here, we use the results for the extrapolation of the Galactic center excess of three compelling analyses, including from the Fermi-LAT Collaboration \cite{Fermi-LAT:2015sau, Fermi-LAT:2017opo}, as well as from Ref.~\cite{Calore:2014xka}, and the recent Ref.~\cite{DiMauro:2021raz}\footnote{We thank Mattia Di Mauro for kindly sharing with us his numerical results.}.

A second excess we will focus our study on is a possible anomaly in the flux of cosmic-ray antiprotons in the 10-20 GeV energy range as measured by the anti-matter spectrometer AMS-02 \cite{AMS:2014bun}, first discovered in Ref.~\cite{Cui:2016ppb, Cuoco:2016eej} and more firmly established in Ref.~\cite{Cholis:2019ejx}. Here as well a lively debate ensued the claim that the excess could be associated with new physics, due to persistent issues with background modeling, source characterization,  modeling of the propagation and of the interaction cross sections responsible for antiproton production and detection (again, we refer the interested Reader to section 5.2 of Ref.~\cite{Slatyer:2021qgc}). Here, we utilize the results of Ref.~\cite{Cholis:2019ejx} when we refer to the ``antiproton excess''.

Finally (albeit, alas, by no means is this a complete list of indirect detection excesses!) here
 we also consider the tantalizing report of potentially several anti-helium ($\Hebar$) events by AMS-02 \cite{Carlson:2014ssa, Coogan:2017pwt} (see also \href{https://indico.cern.ch/event/1275785/}{this recent presentation: https://indico.cern.ch/event/1275785/}). Virtually absent any expected astrophysical background \cite{Carlson:2014ssa}, the reported events could, however, originate from a misidentified matter ${}^{3}{\rm{He}}$ nucleus -- the latter being over $10^{9}$ more frequent; the needed detector simulations that would help establish the mis-identification rate, and assess whether or not it is better than one part in a billion, are currently underway. In view of the possible contamination we entertain, here that only a fraction of the reported candidate events actually consists of $\Hebar$.

 While other studies have addressed some or multiple of the three anomalies described above, here, for the first time, we entertain the possibility that all three be connected with late-forming, small black holes of non-stellar origin (we will refer to them, following \cite{Picker:2023ybp, Picker:2023lup}, as micro-structure black holes (MSBH)). If black holes of non-stellar origin are to be a component, or all, of the cosmological DM, their lifetime must exceed the age of the universe (in fact, the constraints on the lifetime are significantly more stringent, see Ref.~\cite{Korwar:2023kpy} for a recent study). Given that, absent new, beyond the Standard Model, degrees of freedom, the lifetime $\tau$ of a black hole of mass $M$ is
 \begin{equation}
     \tau(M)\simeq t_U\left(\frac{M}{10^{15}\ {\rm g}}\right)^3,
 \end{equation}
 black holes need to be more massive than around $M\gtrsim 10^{15}\ {\rm g}$ (in fact, if black holes are to be most of the cosmological dark matter, their mass must exceed $M\gtrsim3\times 10^{17}\ {\rm g}$, and even if they are only a small fraction, say $10^{-3}$, constraints from the diffuse gamma-ray emission constrain the smallest possible mass $M\gtrsim1\times 10^{17}\ {\rm g}$, see, again, Ref.~\cite{Korwar:2023kpy}).

 Black holes of mass $M$ have a temperature \cite{Hawking:1975vcx}
 \begin{equation}
     T_{\rm BH}=\frac{1}{8\pi G_N M}\simeq 1.06\ {\rm MeV}\ \left(\frac{10^{16}\ {\rm g}}{M}\right),
 \end{equation}
 thus, if black holes are abundant, and primordial, their temperature must be below the MeV. Since Hawking radiation is a quasi-thermal process, the production of particles with energies in the multi-GeV range is exponentially suppressed and essentially absent.

 However, black holes can be much lighter, and hotter, if they were produced in the {\em late universe}, or if they are {\em currently} being produced. Notable possible formation scenarios for such MSBH include, but are not limited to, the collapse of DM structures consisting of particles interacting via Yukawa forces, or the collapse of Q-balls or Fermi balls. Depending on the timescale for collapse, the resulting MSBH can form at different times in the universe, including now. In particular, Ref.~\cite{Kawana:2022lba}, \cite{Hong:2020est} and \cite{Lu:2022paj} consider the collapse of false vacuum remnants in first-order phase transitions arising if particles develop large mass gaps between the two phases and become trapped in the early phase; Ref.~\cite{Savastano:2019zpr} explores a formation scenario where primordial DM halos could be generated during radiation domination by long-range ``fifth-forces'' stronger than gravity; Ref.~\cite{Flores:2022oef} and \cite{Flores:2020drq} study a similar possibility based, however, on a long-range Yukawa force, as does Ref.~\cite{Domenech:2023afs}, albeit in a slightly different realization; Ref.~\cite{Chakraborty:2022mwu} considers instabilities in the density perturbations of a strongly interacting fermion-scalar fluid where the sound speed turns imaginary, leading to exponential growth at sub-Compton scales; finally, Ref.~\cite{Lu:2022jnp} and \cite{Kawana:2021tde} entertain the possible formation of black holes from the collapse of Fermi balls.

 Here, we revisit and expand the idea put forth in \cite{Picker:2023lup}, that MSBH may explain the GeV gamma-ray excess from the Galactic center, and explore the tantalizing possibility that, in fact, such excess can be explained {\em in concert} with the above-mentioned antiproton and anti-helium cosmic-ray anomalies as well. To explore this possibility, we first calculate in detail, in sec.~\ref{sec:bestfit}, the spectrum of gamma rays, antiprotons and anti-helium nuclei expected from MSBH with temperatures in the 10 GeV range, and thus masses around $10^{12}\ {\rm g}$; we then fit the MSBH mass and the rate at which such objects should form today to the three anomalies; in sec.~\ref{sec:predictions} we crucially compute the expected spectrum of other cosmic-ray species, such as neutrinos, positrons, and anti-deuterons, expected if this scenario is realized in nature. Finally, we present our discussion and conclusions in Sec.~\ref{sec:conclusions}.

\section{Best Fit Models for Late-Forming Black Holes}\label{sec:bestfit}

Late-forming black holes are parameterized by the injected black hole mass $m$ and fraction $f_{\rm Myr}$ of the DM energy density collapsing into a black hole per unit time (here, $10^6$ yrs). The injection rate of black holes is given by $dn/dt_{\rm Myr} = (\rho_{\rm{DM}}/m)\ f_{\rm{Myr}} \, $, where $\rho_{\rm DM}$ is the DM energy density and time $t_{\rm Myr} = t/\rm{Myr}$. Thus the total number density of injected BH is $n_{\rm BH} = \tau(m) (\rho_{\rm{DM}}/m)\ f_{\rm{Myr}}$, where $\tau(m)$ is the lifetime of the black hole (in Myr) with mass $m$. Assuming continuous injection of black holes, followed by evaporation with $dM \propto 1/M^{2}$ leads to a ``triangle'' distribution, given by $\phi(M) = 3 M^{2}/m^{3}$ for $M\leq m$ and zero otherwise (see \cite{Picker:2023lup, Picker:2023ybp} for details). Notice that $\int_{0}^{m} \phi(M) dM = 1$. We can rewrite the population of late-forming black holes in terms of the usual $f_{\rm PBH}$ parameter \cite{Carr:2020gox} as 
\begin{equation}
 f_{\rm PBH} = \frac{n_{\rm BH}}{\rho_{\rm DM}} \int_{0}^{m}dM \, M \, \phi(M) = \frac{3}{4} \tau(m) f_{\rm Myr}\, .
\end{equation}
Notice that for $\tau(m)< 1$ the fraction of late-forming black holes in DM is smaller than $f_{\rm Myr}$, that is injected black hole holes which evaporate on time-scales shorter than a million years contribute a much smaller fraction of the  DM than their corresponding injected fraction. This is expected, as the evaporated mass will contribute to the radiation energy density. 

Note that while $m$ and $f_{\rm{Myr}}$ could be constant in the Galaxy with respect to Galactic radius, it is also possible that there is non-trivial DM density dependence on these parameters, such as $f_{\rm {Myr}} = f_{0, \rm{Myr}} (\rho_{\rm{DM}}(r)/\rho_{\rm{DM}}(r_{\oplus}))^{p_f}$ and $m = m_{0} (\rho_{\rm DM}(r)/\rho_{\rm DM}(r_{\oplus}))^{p_m}$. Here $\rho_{\rm DM}(r_{\oplus})$ is the DM energy density at Earth, and $f_{0, \rm{Myr}}, m_{0}$ are independent of radial distance. It has been shown that the choices $p_f=1$ and $p_m=0$ would explain the Galactic center excess morphology~\cite{Picker:2023lup}, and thus we will consider the same parameters for our study. 

Generically, black holes with mass $M$ radiate a ``primary'' particle species $i$ at a rate
\begin{equation}
    \frac{\partial^{2}N_{i}^{\rm{prim}}}{\partial E_{i} \partial t} = \frac{g_{i}}{2\pi} \frac{\Gamma_{i}(T_{\rm{BH}}, E_{i}, s_{i})}{\exp(E_{i}/T_{\rm{BH}})-(-1)^{s_{i}}} \,,
\end{equation}
where $g_{i}$ are degrees of freedom for particle $i$, spin $s_{i}$, energy $E_{i}$, and species dependent gray-body factors $\Gamma_{i}$ \cite{Carr:2020gox}. %The black hole temperature $T_{\rm{BH}}$ is given by
%\begin{equation}
%    T_{\rm{BH}} = \frac{1}{8\pi G_{N} M} \approx 10 \, \rm{GeV} \, \Big{(}\frac{10^{12} \, \rm{g}}{M}\Big{)}\,.
%\end{equation}
In addition to the primary emission $\partial^{2} N^{\rm prim}_{i}/\partial E_{i} \partial t$, we also have to take into account the secondary emission $\partial^{2} N^{\rm secondary}_{i}/\partial E_{i} \partial t$ from hadronization and decays of unstable states. We utilize the output primary and secondary emission rates from the state-of-the-art {\tt BlackHawk} code~\cite{Arbey:2019mbc}. The total emission rate for stable species $i$, $\partial^{2} N_{i}/\partial E_{i} \partial t$ is given by the addition of primary and secondary emission rates for that species.

\begin{figure*}[t]
    \includegraphics[width=0.8\textwidth]{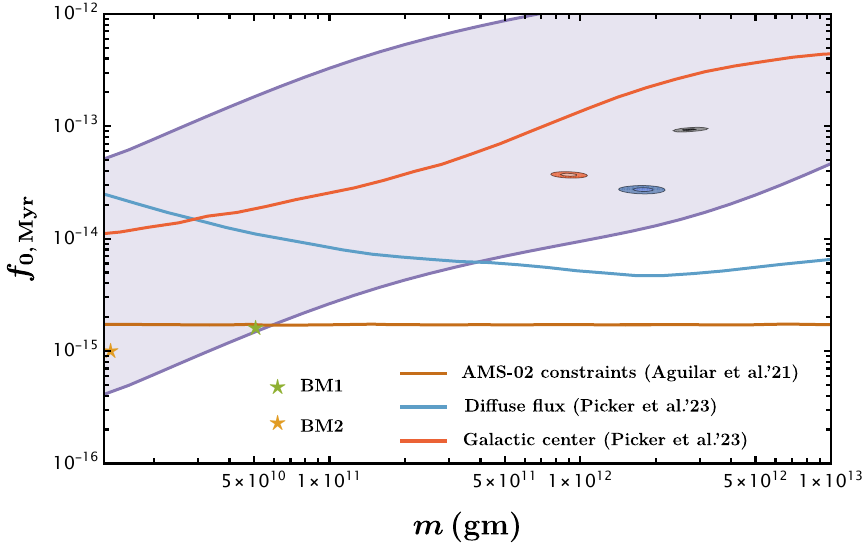} 
    \caption{Best fit regions for explaining the Galactic center excess (red, blue, and black contours) and $\Hebar$ detection (purple contour) in the parameter space of injected BH mass $m$ and injected BH fraction $f_{0, \rm{Myr}}$. The red~\cite{Fermi-LAT:2015sau, Fermi-LAT:2017opo}, blue~\cite{Calore:2014xka}, and black~\cite{DiMauro:2021raz} contours represent analyses done with different background-subtracted models for the Galactic center excess. The brown curve shows the 95\% CL bound obtained using AMS-02 data for the antiproton-to-proton flux ratio~\cite{AMS:2021nhj}. For the anti-helium excess analysis, we consider the detection of 1 event in 10 years at a kinetic energy of $K=12 \, \rm{GeV/n}$~\cite{Coogan:2017pwt}. The green star \((5 \times 10^{10} \text{g}, 1.5 \times 10^{-15})\) and orange star \((10^{10} \text{g},  10^{-15})\) represent our benchmark points, which could potentially explain the $\Hebar$ detection and are compatible with the strongest antiproton constraint. We also show the current best constraints on late-forming black holes from diffuse extra-galactic gamma-ray and x-ray flux (blue curve) and gamma rays from the Galactic center (red curve)~\cite{Picker:2023ybp}. Note that the diffuse background constraints are model-dependent and assume BH injection at higher redshift~\cite{Picker:2023lup, Picker:2023ybp}. }
    \label{fig:bestfit}
\end{figure*}

Late-forming MSBHs in the Galactic center radiate photons, and the expected gamma-ray flux is 
\begin{widetext}
    \begin{equation}
 E_{\gamma}^{2}\frac{d\Phi_{\gamma}}{dE_{\gamma}} = \tau(m) \Big{(}E_{\gamma}^{2}\int_{0}^{m}dM\,\phi(M) \frac{d^{2}N_{\gamma}}{dE_{\gamma} dt} \Big{)}\, \Big{(}\int_{\Delta \Omega} \frac{d\Omega}{4\pi} \int_{\rm{l.o.s}}dr \frac{\rho_{DM}(r)}{m} f_{\rm{Myr}}(r)\Big{)}\, ,
\end{equation}
\end{widetext}

where $\Delta \Omega$ for region of interest. For DM density distribution we assume a NFW profile \cite{Navarro:1995iw}, 
\begin{equation}
    \rho_{\rm DM}(r) =\frac{\rho_{0}r_{s}^{3}}{r_{s}(r_{s}+r)^{2}},
\end{equation} with $r_{s}=25\,\rm{kpc}$, distance to Earth $d=8.33 \, \rm{kpc}$, and energy density at Earth $\rho_{\rm DM}(r_{\oplus}) = 0.3 \, \rm{GeV}\rm{cm}^{-3}$. To find the parameter space in $m, f_{0, \rm Myr}$ that would explain the Galactic center excess, we use three different datasets each with different background modeling. The dataset from \cite{Fermi-LAT:2015sau, Fermi-LAT:2017opo} considers region of $2^{\circ}\leq |l| \leq 10^{\circ}$, $2^{\circ}\leq |b| \leq 10^{\circ}$ towards Galactic center for their analysis, whereas \cite{Calore:2014xka} uses $|l|\leq 20^{\circ}, 2^{\circ}\leq |b| \leq 20^{\circ}$ and \cite{DiMauro:2021raz} considers $|l|, |b|\leq 40^{\circ}$. In Fig.~\ref{fig:bestfit} We show $68\%, 95\%$ confidence interval for these datasets using red, blue, and black contours respectively. 

The emission of quarks and gluons from MSBHs also leads to hadronization and the production of hadrons, particularly antiprotons. The flux of antiprotons at Earth depends on both the emission rate and subsequent processes of diffusion and energy loss. We use the publicly available version of the \texttt{DRAGON2} code\footnote{We use \url{https://github.com/cosmicrays/DRAGON2-Beta_version}, particularly its latest version \url{https://github.com/tospines/Customised-DRAGON2_Antinuclei}} to perform particle propagation for both the background processes contributing to the antiproton flux and the contribution from MSBH evaporation~\cite{Evoli:2016xgn, Evoli:2017vim}. We fit the AMS-02 data to obtain bounds on the parameter space of MSBHs. The antiproton flux from MSBHs has a source term:
\begin{equation}\label{eq:Qpbar}
    Q_{\bar{p}}^{\text{MSBH}} = \frac{3}{4} \tau_{\rm{Myr}}(m) \frac{f_{0, \rm{Myr}}}{m} \frac{\rho_{\rm{DM}}^{1+p_{f}}}{\rho_{\rm{DM}}(r_{\odot})^{p_{f}}} \int_{0}^{m}dM \phi(M) \frac{d^{2}N_{\bar{p}}}{dE dt},
\end{equation}
where $\frac{d^{2}N_{\bar{p}}}{dE dt}$ is the rate of antiproton emission from an evaporating black hole.

The observed antiproton flux is a combination of contributions from background cosmic-ray sources and the MSBH source term. While the contribution from the MSBH source term depends on the parameters of our model, there is uncertainty regarding the antiproton production source term from the background sources, particularly due to the antiproton production cross-section. To account for this uncertainty, we follow the treatment of Ref.~\cite{Cholis:2019ejx} and consider an energy-dependent scaling factor:
\begin{equation} \label{eq:scalingeqn}
    N_{\rm{CS}}(K) = a + b \log(K/\rm{GeV}) + c \log(K/\rm{GeV})^{2},
\end{equation}
where $K = E - m_{p}$ is the kinetic energy of the antiprotons. As antiprotons (or any charged particle) enter the solar system, their flux is modulated by the solar wind and the Sun's magnetic field. The effect of this modulation on the observed flux at Earth is given by~\cite{1968ApJ...154.1011G}:
\begin{equation}
    \frac{d\Phi_{\bar{p}}(K_{\oplus}, r_{\oplus})}{dK_{\oplus}} = \frac{p_{\oplus}^{2}}{p^{2}} \frac{d\Phi_{\bar{p}}(K, r_{\odot})}{dK}, \quad K=K_{\oplus} + |Ze|\phi_{F}^{\pm},
\end{equation}
where $p=\sqrt{2 m_{p} K + K^{2}}$ is the momentum, $|Ze|$ is the electric charge of the particle (here $Z=1$), and $K_{\oplus}, p_{\oplus}$ are the kinetic energy and momentum of the particle after antiprotons reach Earth. The $\phi_{F}^{\pm}$ is the Fisk potential, which depends on the charge of the particle. We parameterize this potential as~\cite{Reinert:2017aga}:
\begin{equation}\label{eq:solmod}
    \phi_{F}^{\pm} = \phi_{0} \pm \phi_{1}^{\pm} F(R/R_{0}),
\end{equation}
where $F(R/R_{0}) = R_{0}/R$ with $R_{0}=1 \, \rm{GV}$ is the reference rigidity and $R$ is the rigidity of antiprotons. We take $\phi_{0} = 0.58 \, \rm{GV}$, and accounting for the solar polarity phase during the AMS-02 data-taking period, we take $\phi_{1}^{+} = 0$ and allow for a non-vanishing $\phi_{1}^{-}$~\cite{DelaTorreLuque:2024ozf}. 

We fit a six-parameter model, with three parameters \(a\), \(b\), and \(c\) for the scaling of antiproton cross sections, one parameter \(\phi_{1}^{-}\) for the solar modulation potential, and two parameters \(m\) and \(f_{0, \rm{Myr}}\) for MSBH to the latest AMS-02 data for the antiproton-to-proton flux ratio~\cite{AMS:2021nhj}. The results of the MCMC simulation are shown in Appendix \ref{sec:appendix}. After marginalizing over the cross-section scaling and the solar modulation parameter, the 95\% confidence level (CL) bound obtained in the \(f_{0, \rm{Myr}}-m\) plane is shown by the brown curve in Fig. \ref{fig:bestfit}.

In addition to protons and neutrons from hadronization, higher atomic elements could also be produced from BH evaporation. In particular, in view of the reported tentative AMS-02 events, here we focus on $\Hebar$. The production spectra of $\Hebar$ from an elementary particle $i$ which undergoes hadronization to produce nucleons $p, n$, can be estimated using the coalescence model \cite{Ibarra:2012cc, Carlson:2014ssa}
\begin{equation}\label{eq:coalascence}
    \frac{dN^{i}_{\Hebar}}{dE_{\Hebar}} = \frac{m_{\Hebar}}{m_{p}^{2} \, m_{n}} 3 \Big{(}\frac{p_{0}^{3}}{8 p_{\Hebar}}\Big{)}^{2} \Big{(}\frac{dN^{i}_{\bar{p}}}{dE_{\bar{p}}}\Big{)}^{2}  \frac{dN^{i}_{\bar{n}}}{dE_{\bar{n}}}\, ,
\end{equation}
where $p_{0}=0.246 \pm 0.038 \, \rm{GeV}$ is coalescence momentum for $\Hebar$ production~\cite{Coogan:2017pwt}, and $E_{\Hebar}, p_{\Hebar}, m_{\Hebar}$ are, respectively, the energy, momentum and mass of the $\Hebar$ nucleus. Here,  $dN_{\bar{p}/ \bar{n}}/dE_{\bar{p}/\bar{n}}$ are the antiproton and antineutron spectra at production (thus prior to diffusion and energy losses) from hadronization. The net production rate of $\Hebar$ is then the convolution of the production rate of the elementary particle $i$ with the spectra of $\Hebar$ evaluated from the coalescence equation of Eq.~(\ref{eq:coalascence}) i.e.
\begin{equation}
    \frac{d^{2}N_{\Hebar}}{dE_{\Hebar} dt} = \sum_{i}\int dE_{i}' \frac{d^{2}N_{i}}{dE_{i}' dt} \frac{dN^{i}_{\Hebar}(E_{\Hebar}, E_{i}')}{dE_{\Hebar}} \,,
\end{equation}
where the variable of integration $E_{i}'$ is the energy of elementary particle $i$. In addition to the production rate, again, one needs to take into account diffusion, transport, and solar modulation in order to estimate the flux at the Earth surface. We implement all of these utilizing the assumptions and tools provided and detailed upon in Ref.~\cite{Cirelli:2010xx}. The purple region in Fig.~\ref{fig:bestfit} shows the parameter space explaining 1 event in 10 years at $K_{\Hebar} = 12 \, \rm{GeV/n}$. The large error bars are a result of huge uncertainty on the coalescence momentum, the transport properties, and the lack of data.  Scaling for a larger number of events simply and trivially shifts the contours to larger values of $f_{\rm 0,Myr}$.

In addition to the contours explaining various excess, we also show current best {\em constraints} on this parameter space coming from gamma rays from the Galactic center (red curve) and diffuse gamma-ray and x-ray flux (blue curve)~\cite{Picker:2023lup}. The diffuse flux constraints here assume the formation of late-forming MSBHs at much higher redshift and thus are model-dependent. 

Our benchmarks are $(5\times 10^{10} \rm{g}, 1.5 \times 10^{-15})$ for green star, and $(10^{10} \rm{g}, 10^{-15})$ for orange star. The lifetime for green star (BM1) is $5.6 \times 10^{4}\, \rm{s}$ and spawning rate of 3 BHs per cubic parsec per Myr, while for orange star (BM2) we get lifetime of $450 \, \rm{s}$ and spawning rate of 10 BH per cubic parsec per Myr. We use this benchmark points for our predictions for other indirect detection signals, with different particle messengers,  in the following Sec.\ref{sec:predictions}.

\section{Model Predictions: antideuterons, positrons, neutrinos, and gamma rays}\label{sec:predictions}

\begin{figure*}[htbp]
\centering
\subfloat[Antideuteron flux]{%
  \includegraphics[width=0.48\textwidth]{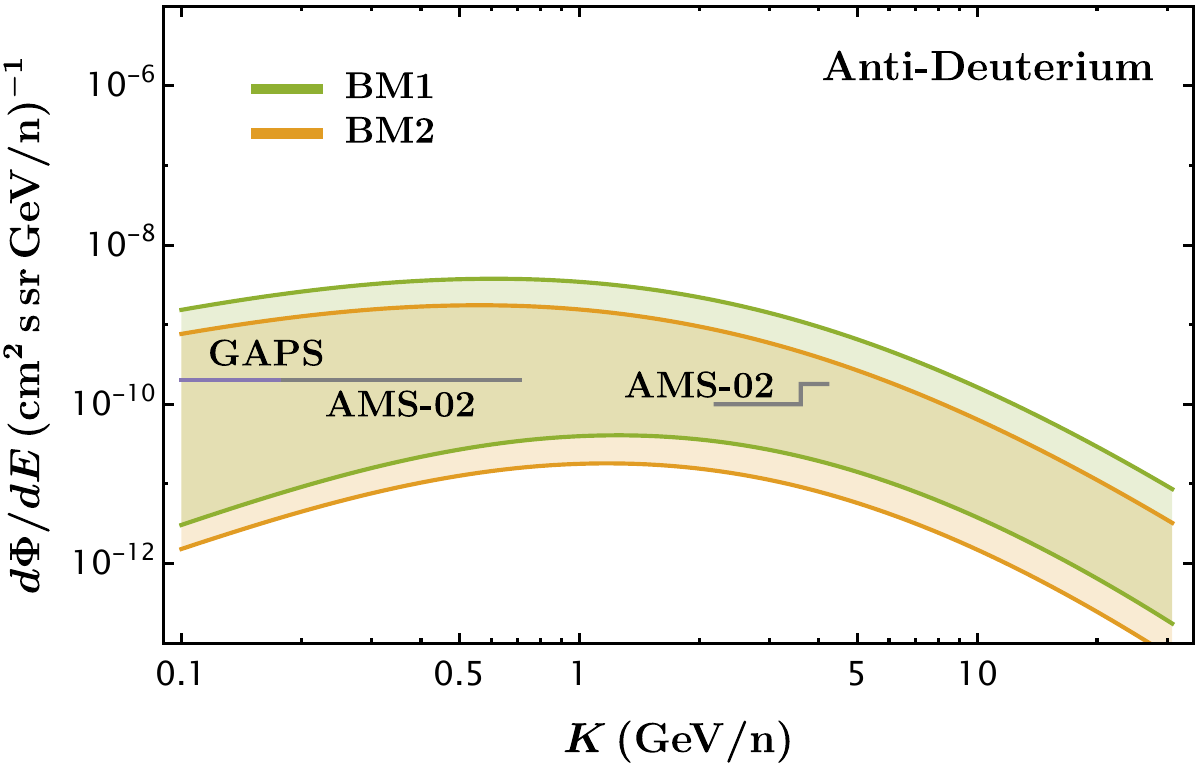}%
}
\hfill
\subfloat[Antiproton fraction]{%
  \includegraphics[width=0.48\textwidth]{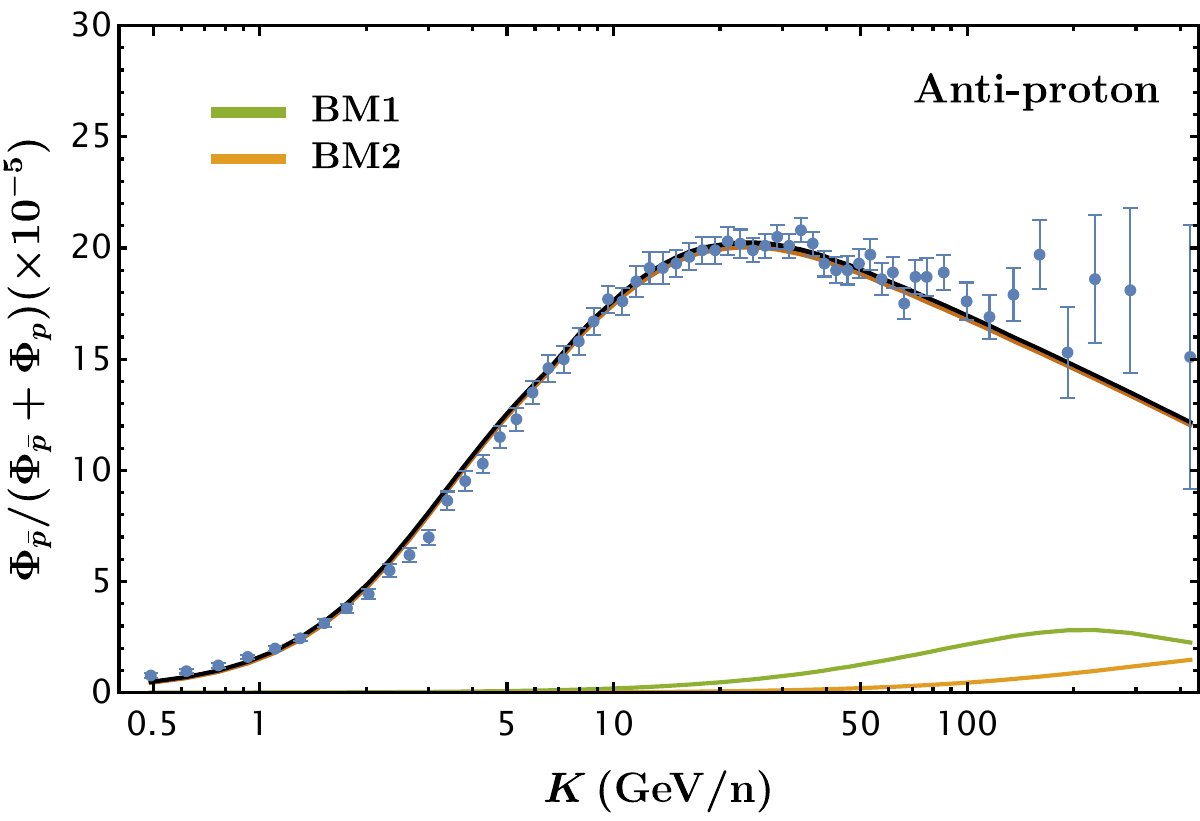}%
}

\vspace{0.1cm}

\subfloat[Neutrino flux]{%
  \includegraphics[width=0.48\textwidth]{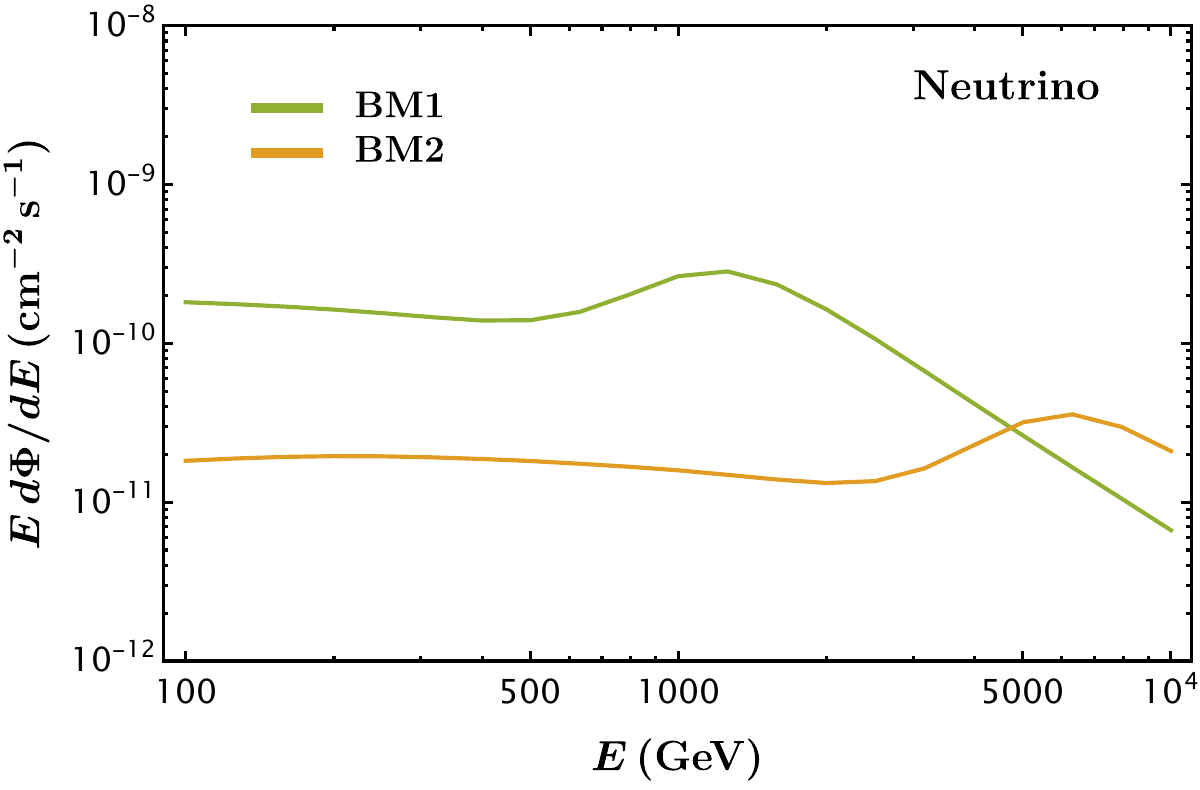}%
}
\hfill
\subfloat[Gamma-ray flux]{%
  \includegraphics[width=0.48\textwidth]{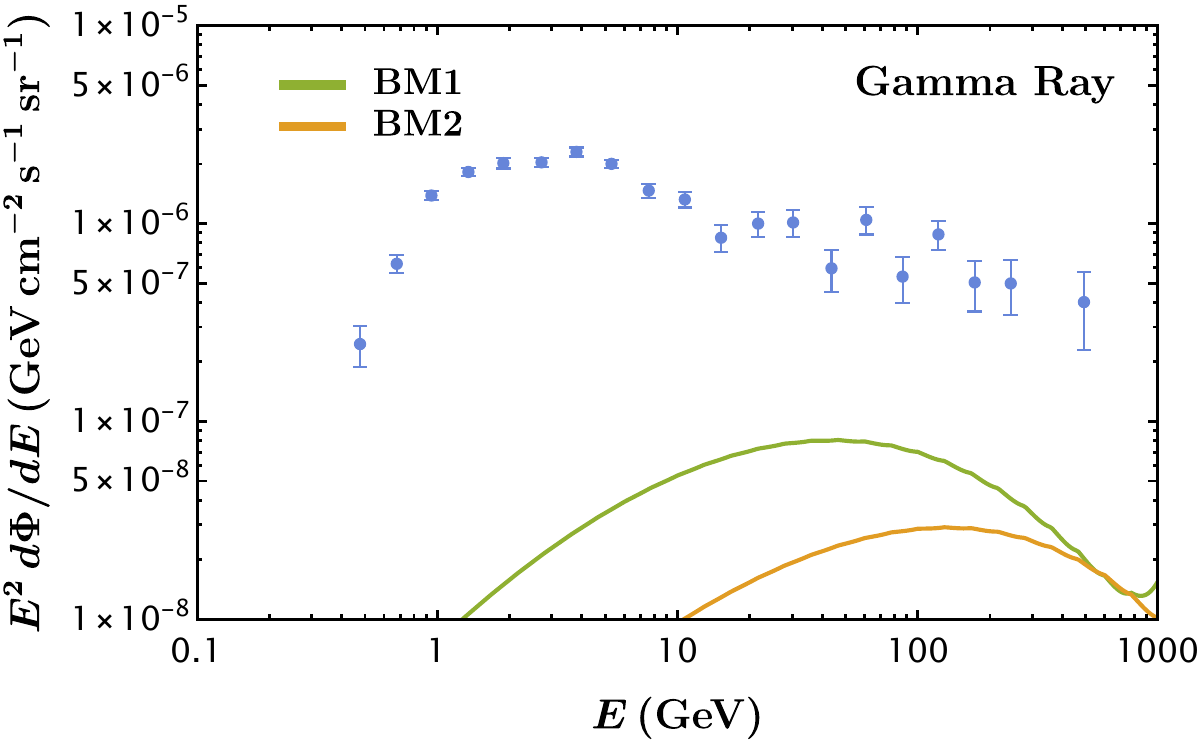}%
}

\vspace{0.1cm}

\subfloat[Positron ratio]{%
  \includegraphics[width=0.48\textwidth]{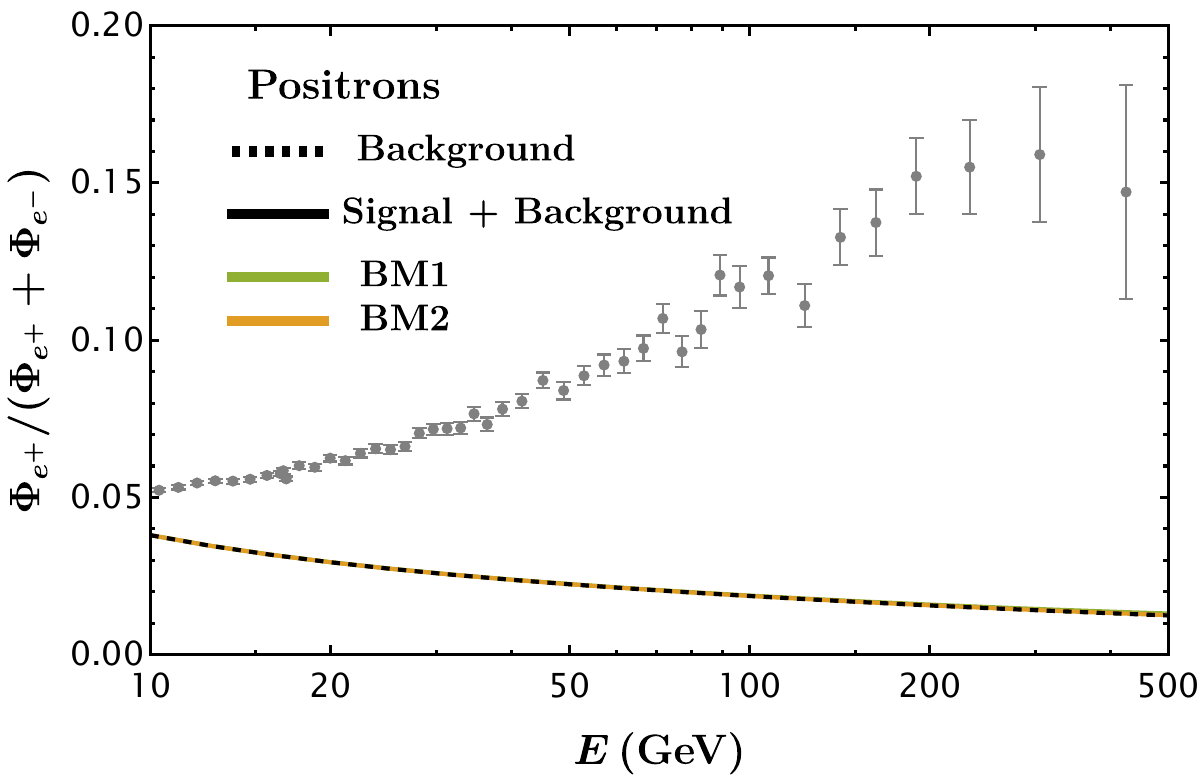}%
}
\hfill
\subfloat[Antihelium flux]{%
  \includegraphics[width=0.48\textwidth]{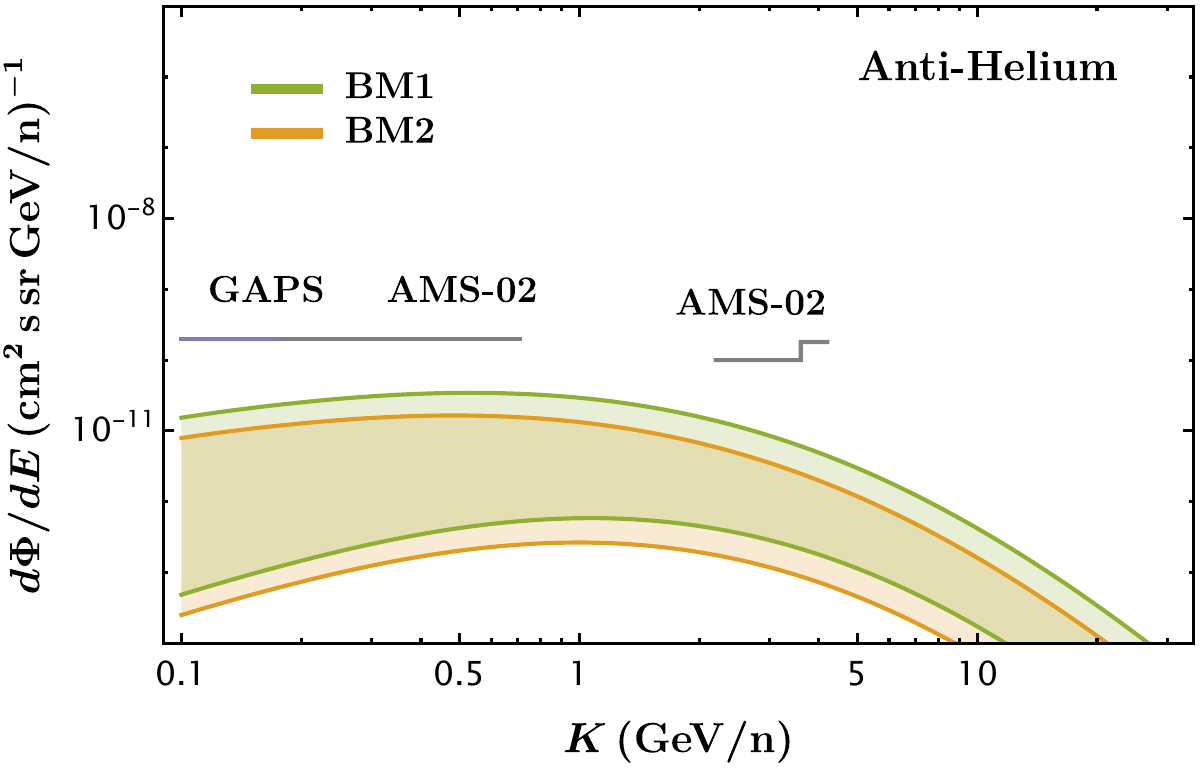}%
}

\caption{Prediction for the antideuteron flux, antiproton faction, neutrino flux,  gamma-ray flux from the Galactic center, positron ratio, and antihelium flux for the benchmark points BM1 and BM2 shown as a green and orange star star in Fig.\ref{fig:bestfit}. For antideuterons, in the top-left panel, we show the sensitivity limits for GAPS (purple) and AMS-02 (gray) experiments~\cite{Aramaki:2015laa, Aramaki:2015pii}. The top right panel shows the antiprotons fraction, with the observed data points shown with the error bars~\cite{AMS:2016oqu}. In the middle left panel, we show the flavor averaged neutrino flux. The middle right panel compares the expected gamma-ray spectrum for the benchmark point with the observed GCE excess, where the data points with error bars are taken from the analysis done in \cite{Fermi-LAT:2015sau, Fermi-LAT:2017opo}. In the bottom left panel, we compare the expected positron fraction from our model with the observed data points taken from~\cite{AMS:2014bun}, and we take the background model from~\cite{Baltz:1998xv, Cirelli:2008id}. In the bottom right panel, we comapre the antihelium expected spectra with sensitivity of GAPS (purple) and AMS-02 (gray) experiments.}
\label{fig:prediction}
\end{figure*}

In this section we use our benchmark points discussed above to estimate the spectrum and expected number of events for positrons, antideuterons ($\bar{D}$s), and neutrinos. We will also comment on the positron excess. 

Similar to the case of $\Hebar$ discussed above, the antideuteron production from evaporating BHs (first computed in Ref.~\cite{Barrau:2002mc}) is estimated using  a coalescence model, here
\begin{align}
    \frac{dN^{i}_{\Dbar}}{dE_{\Dbar}} &= \frac{m_{\Dbar}}{m_{\bar{p}} \, m_{\bar{n}}} \Big{(}\frac{p_{0}^{2}}{6 \, p_{\Dbar}}\Big{)}  \frac{dN^{i}_{\bar{p}}}{dE_{\bar{p}}}  \frac{dN^{i}_{\bar{n}}}{dE_{\bar{n}}}\, , \\ \nonumber \\
     \frac{d^{2}N_{\Dbar}}{dE_{\Dbar} dt} &= \sum_{i}\int dE_{i}' \frac{d^{2}N_{i}}{dE_{i}' dt} \frac{dN^{i}_{\Dbar}(E_{\Dbar}, E_{i}')}{dE_{\Dbar}} \,.
\end{align}
For the coalescence momentum we adopt $p_{0}=0.192 \pm 0.030 \, \rm GeV$~\cite{Coogan:2017pwt}. In Fig.~\ref{fig:prediction} top left panel, we show predictions for the $\Dbar$ spectra for our benchmark points, taking into
account the uncertainties in coalescence momentum and transport properties. We also show the sensitivity limits of AMS-02 and GAPS experiments for $\Dbar$ detection~\cite{2008ICRC....4..765C, Aramaki:2015laa, Aramaki:2015pii} with purple and gray lines, respectively. We deduce that a significant antideuteron flux is expected both at AMS-02 and GAPS,  for both of our benchmark points.

In the middle left panel of Fig.\ref{fig:prediction}, we show the expected spectra for neutrinos, averaging over flavors, from the Galactic center from the region $|l|, |b|< 10^{\circ}$. Using the effective area  $\sim 100 \, \rm{cm}^{2}$ for a $1000 \, \rm GeV$ neutrino~\cite{2010arXiv1003.5715K}, and runtime of $10 \, \rm yrs\approx 10^{8} \rm{s}$, we expect $\mathcal{O}(10)$ events in IceCube detector, significantly smaller compared to the astrophysical backgrounds from atmospheric neutrinos \cite{2010arXiv1003.5715K}. 

In the bottom panel of Fig.\ref{fig:prediction}, we show the expected spectra for positrons for our benchmark point. For diffusion and energy loss equations, we again rely on the analysis from \cite{Cirelli:2010xx}. For the background model, we use the analytical forms of \cite{Baltz:1998xv, Cirelli:2008id}. We can see that the predicted signal is significantly below the observed positron excess \cite{AMS:2014bun}. 

In the top right panel, we show the comparison of the observed antiproton {\em fraction} \cite{AMS:2016oqu}, with what is expected from our benchmark model. In the middle right panel, we show the comparison of gamma-ray flux from the Galactic center from our benchmark model with the background-subtracted data from the Fermi Collaboration analysis \cite{Fermi-LAT:2015sau, Fermi-LAT:2017opo}. Finally, in the bottom right panel we show expected anti-helium flux from MSBH evaporation for our benchmark points taking into account the uncertainty in the coalescence momenta, along with the detection sensitivities of GAPS and AMS-02 experiment~\cite{2008ICRC....4..765C, Aramaki:2015laa, Aramaki:2015pii}.

\section{Discussion and Conclusions}\label{sec:conclusions}
  
We have investigated the intriguing possibility that three long-standing anomalies—the Galactic center gamma-ray excess, the antiproton excess, and the tentative anti-helium-3 events reported by the AMS-02 Collaboration—may all originate from late-spawned, light black holes of non-stellar origin. Utilizing the latest particle propagation codes and Markov Chain Monte Carlo techniques, we find no antiproton excess and establish the leading constraint on light, late-spawning black holes. Consequently, we exclude the possibility of explaining the gamma-ray excess with late-forming black holes. However, the anti-helium-3 events could still be explained by late-forming black holes of smaller mass, around $10^{10}$ g, injected at a rate of approximately 10 black holes per cubic parsec per mega-year.

To cross-constrain this scenario, we explored the implications for orthogonal astrophysical messengers, such as antideuterons, positrons, and neutrinos, which also necessarily arise from the evaporation of the late-forming black holes. We found that an antideuteron signal is likely to appear if this scenario is realized in nature, detectable by both AMS-02 and GAPS. The predictions for the high-energy neutrino flux are far below the atmospheric neutrino background, and the positron flux is also small compared to the observed high-energy cosmic-ray positron flux.

In conclusion, using the observed antiproton-to-proton ratio, we obtain strong limits on late-forming black holes. We find that small mass black holes could still explain the observed anti-helium-3 events within a specific mass range. This scenario leads to concrete predictions for the spectra of multiple messengers. The advent of GAPS and future measurements of the antideuteron flux with AMS-02 will conclusively test the scenario under consideration here.

\acknowledgments
We thank Mattia Di Mauro and Philip Von Doetinchem. S.P. is partly supported by the U.S. Department of Energy grant number de-sc0010107. M.K. acknowledges support from the National Science Foundation (Grant No. PHY-2020275).

\appendix
\section{MCMC results}\label{sec:appendix}
\begin{figure*}
    \centering
    \includegraphics[width=0.78\textwidth]{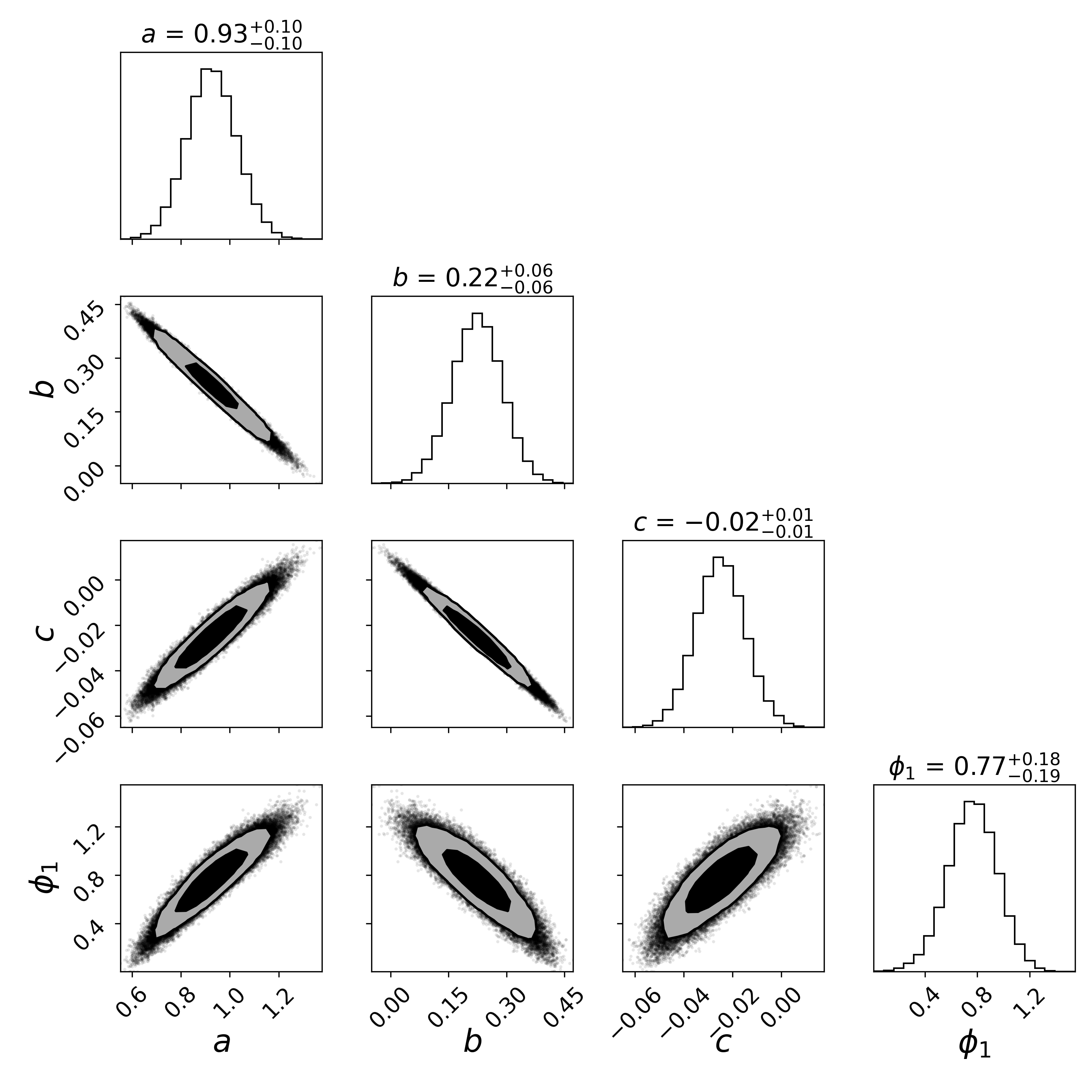} 
    \caption{The PDFs of the parameters used in our model. The black and gray contours show 68\% and 95\% credible regions.}
    \label{fig:mcmc}
\end{figure*}
In this appendix, we show the probability distribution functions (PDFs) of the parameters mentioned in the main text. In Fig.~\ref{fig:mcmc}, we show the 68\% and 95\% credible regions for the cross-section scaling parameters ($a, b, c$) and the solar modulation parameter $\phi_{1}^{-}$, along with their mean values and standard deviation.

\setlength{\bibsep}{3pt}
\bibliographystyle{apsrev4-2}
\bibliography{sample}

\end{document}